\documentclass[twocolumn,letterpaper,nofootinbib,prd,amsmath]{revtex4}

\usepackage[dvips]{graphics,color}
\usepackage{epsfig}
\usepackage{hyperref}
\newcommand{\RR}{\hbox{$I$\kern-3.8pt $R$}}
\newcommand{\lessthansimilarto}{\lower3pt\hbox{$\buildrel{<}\over{\sim}$}}

\begin{document}

\title{Multigrid elliptic equation solver with adaptive mesh refinement}  
\author{J.~David Brown}
\affiliation{Department of Physics, North Carolina State University,
Raleigh, NC 27695 USA}
\author{Lisa L.~Lowe}
\affiliation{Department of Physics, North Carolina State University,
Raleigh, NC 27695 USA}

\begin{abstract}
In this paper we describe in detail the computational algorithm used by our parallel multigrid 
elliptic equation solver with adaptive mesh refinement. Our code uses truncation error estimates 
to adaptively refine the grid as part of the solution process. The presentation includes a discussion of 
the orders of accuracy that we use for 
prolongation and restriction operators to ensure second order accurate results and to minimize computational 
work. Code tests are presented that confirm the overall second order accuracy  and 
demonstrate the savings in computational resources provided by adaptive mesh refinement.
\end{abstract}

\maketitle

\section{Introduction}
Elliptic equations appear throughout engineering, science, and mathematics. Our primary 
interest is in elliptic problems that arise in the context of numerical relativity. Currently, the field of numerical 
relativity is being driven by rapid progress on the experimental front. There are several ground--based gravitational
wave detectors in operation today, and their sensitivities are quickly approaching a level at which interesting 
science can be done. There are also plans for a space--based gravitational wave detector, LISA, to be launched around 
2012. The scientific payoff of these instruments will depend largely on our ability to theoretically predict and explain  
the observed signals. For both ground--based and space--based gravitational wave detectors, the most 
common and strongest signals are expected to come from colliding black holes. Thus, much of the numerical 
relativity community has directed its efforts toward modeling binary black hole systems.

When black holes spiral together and collide, they generate  gravitational waves. 
The black hole ``source'' region has a  length scale of $~GM/c^2$, where $G$ is Newton's constant, $M$ is the 
total mass of the two black holes, and $c$ is the speed of light. The gravitational waves produced by the 
source have a length scale up to $\sim 100\,GM/c^2$. Herein lies one of the challenges of modeling binary black hole 
systems with finite difference methods. 
The source region requires grid zones of size $\lessthansimilarto\, 0.01\,GM/c^2$ to accurately 
capture the details of the black holes' 
interaction, while the extent of the grid needs to be several hundred $GM/c^2$ to accurately capture the 
details of the gravitational wave signal. Many research groups 
in numerical relativity are starting to use adaptive mesh refinement (AMR) techniques to deal with this 
discrepancy in length scales 
[1--11].
\nocite{Choptuik:1992jv,Brugmann:1996kz,Brugmann:1997uc,Papadopoulos:1998dk,Diener:1999mf,New:2000dz,Choptuik:2003ac,Choi:2003ba,Schnetter:2003rb,Bruegmann:2003aw,Imbiriba:2004tp} 
With AMR the grid resolution is allowed to vary across the computational 
domain so that computational resources can be concentrated where they are most needed. For the binary 
black hole problem, we need a high resolution region to cover the small--scale detail of the source, but the 
gravitational waves far from the source can be modeled with sufficient accuracy 
using a much lower resolution grid.  

Elliptic equations occur in several contexts in numerical relativity. Einstein's theory of gravity is a system of 
partial differential equations consisting of four constraint equations and a set of evolution equations 
(see for example Ref.~\cite{MTW}). 
The constraint equations restrict the data at each time step so in particular the initial data cannot be chosen 
freely. With suitable assumptions about the nature of the initial data, the constraint 
equations can be written as an elliptic system \cite{Cook:2000vr}. 

Having solved the constraints for the initial data, 
those data are evolved forward in time by the evolution equations. At an analytical level, the evolution equations 
guarantee that the constraint equations will continue to be satisfied. However, in numerical modeling, numerical 
errors will introduce violations of the constraints. These violations can be disastrous because the evolution 
equations admit unphysical, constraint violating solutions that grow exponentially [14--17].
\nocite{Kidder:2001tz,Lindblom:2002et,Scheel:2002yj,Lindblom:2004gd} 
One possible strategy for 
preventing this disaster is to impose the constraints during the evolution, which means solving the elliptic 
constraint equations after each time step [18--20].
\nocite{Choptuik:2003as,Anderson:2003dz,Matzner:2004uu}

Elliptic equations also arise in numerical relativity when one is faced with choosing a coordinate system. 
In Einstein's theory the coordinate system must be chosen dynamically as the gravitational field evolves forward in time. 
The choice of coordinate system can have a dramatic effect on the performance of a numerical relativity code. 
Researchers have developed many different strategies  
for choosing a coordinate system. Some of these strategies require the solution of  elliptic, parabolic or hyperbolic 
equations, and some involve algebraic conditions. Some researchers feel that the elliptic 
conditions might be best, but the cost of solving elliptic equations at each time step has made the other choices 
more practical and more popular. 

In the numerical relativity community we need the capability of solving elliptic equations  quickly on 
adaptive, non--uniform grids. No doubt this same need exists in other areas of science and applied mathematics. 

Multigrid methods originated in the 1960's with the work of Fedorenko and Bakhvalov 
\cite{Fedorenko:1962,Fedorenko:1964,Bakhvalov:1966}. They were further developed 
in the 1970's by Brandt \cite{BrandtA:1977,Brandt77}, and are now the preferred 
methods for solving elliptic 
partial differential equations. The advantage of multigrid is its speed---multigrid algorithms only require 
order $N^3$ operations to solve an elliptic equation, where $N^3$ is the number of grid points. In this paper we describe 
our code, AMRMG, which solves nonlinear elliptic equations using multigrid methods with adaptive mesh 
refinement. The idea of combining multigrid with AMR is not new [21--24],
\nocite{Brandt77,BrandtA:1977,Martin:1996,MGTutorial}
although there are a number of features of our code that distinguish it from  the discussions we have 
seen. In particular, AMRMG uses  cell--centered data, as opposed to node centered data. AMRMG uses the Full 
Approximation Storage (FAS) algorithm, and therefore is not restricted to linear elliptic equations.  AMRMG uses 
the Paramesh package to implement parallelization and to organize the multigrid structure \cite{MacNeice00,parameshmanual}. 
In Ref.~\cite{Brown:2004km} we used AMRMG to solve numerically for distorted black hole initial data. 

AMRMG is currently set up to solve second order equations that are
semi--linear (the second order derivative terms are linear in the unknown
field). The FAS scheme is applicable for fully nonlinear equations as
well, and in principle AMRMG can be modified to solve any nonlinear
equation. The equations of current interest for us are semi--linear,
therefore we have not tested AMRMG on any fully nonlinear systems.

In this paper we describe the algorithm behind AMRMG in detail. In Sec.~\ref{sec2} we present the overall conceptual 
framework behind our code, and discuss some of the choices made in its development. Section \ref{sec3} is devoted to 
a discussion of guard cell filling, which determines the coupling between fine and coarse grid regions. 
In Sec.~\ref{sec4} we review the 
FAS algorithm and in Sec.~\ref{sec5} we discuss in detail the restriction and prolongation operators used by AMRMG. 
In Sec.~\ref{sec6} we describe the calculation of the relative truncation error and how it is used to control the grid structure. 
Section \ref{sec7} contains the results of a number of code tests involving the calculation of initial data for  
numerical relativity. 
\section{Multigrid with AMR}\label{sec2}
The simplest technique for solving an elliptic problem is relaxation. 
The equation (or system of equations) 
is written in discrete form as $f_i(\phi) = g_i(\phi)$ where $i$ labels the grid points and $\phi_i$ denotes 
the numerical solution.  A ``relaxation 
sweep'' consists of refining the approximate solution $\phi_i^{\rm old}$ by solving the system 
$f_i(\phi^{\rm new}) = g_i(\phi^{\rm old})$ for $\phi_i^{\rm new}$. 
[In the simplest case $f$ is the identity and 
relaxation is written as $\phi^{\rm new}_i = g_i(\phi^{\rm old})$.]
The success of the relaxation method 
depends on how the finite difference equations are split into a left--hand side $f_i(\phi)$ and a right--hand 
side $g_i(\phi)$. When relaxation does work, it is slow to converge. In particular the long wavelength features 
of the solution must slowly ``diffuse'' across the grid with successive relaxation sweeps.  

To solve a problem with multigrid methods we introduce a hierarchy of grids with different resolutions. 
For the moment, consider the case in which we seek the numerical solution of an elliptic equation 
on a uniform grid of size $N^3$ 
that covers the entire computational domain. We introduce grids of size $(N/2)^3$, $(N/4)^3$, {\it etc.}, each 
covering the computational domain. On each grid the  finite difference equation, or an associated equation, 
is solved by relaxation. The equations to be solved on each multigrid level are discussed in Sec.~\ref{sec4}. 
For now, we simply note that the equations are chosen so that relaxation on the coarse grids quickly
captures the long wavelength features of the solution. Relaxation on the fine grids captures the 
short wavelength features. The grids in the multigrid hierarchy communicate with one another 
through restriction and prolongation operators. Restriction 
takes data on a grid in the hierarchy and restricts it to the next coarsest grid. Prolongation 
takes data on a grid in the hierarchy and interpolates it onto the next finest grid.
Different multigrid algorithms use different sequences of grids in solving elliptic problems, but
the most basic sequence is the V-cycle. In a  multigrid V-cycle one starts with the finest grid, 
steps down the grid hierarchy to the coarsest grid, then steps back up to the finest grid. 

In the context of a time--dependent problem, adaptive mesh refinement (AMR) means that the grid structure adapts in time 
to meet the changing demands as the fields evolve. In the context of an elliptic (time--independent) problem, AMR 
means that the grid structure is determined adaptively, as part of the solution process, in an attempt to minimize 
numerical errors. 

We use the Paramesh package to organize the grid structure for our code. Paramesh covers the 
computational domain with blocks of data of varying spatial resolution. These blocks form a tree data--structure. 
They are logically Cartesian, consisting of a fixed number of cells. We typically use $8^3$ cells for each 
block. Figure  1 shows an example one--dimensional grid. 
\begin{figure*}[htb]
\includegraphics{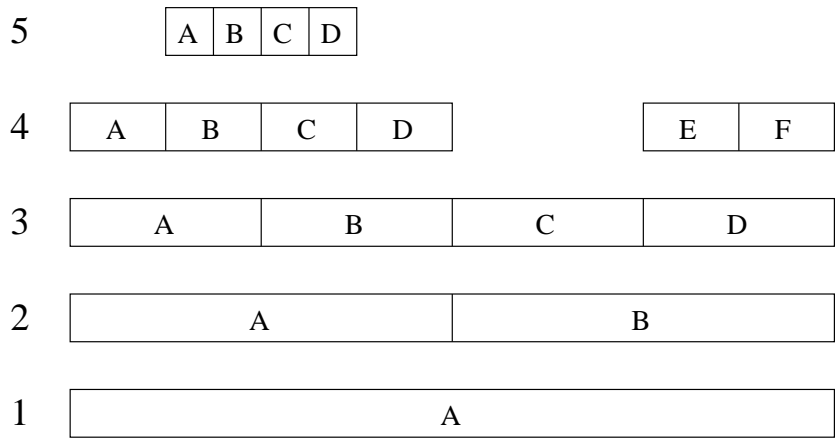}
\caption{Example of a one--dimensional grid structure. The numbers on the left denote the resolution level, and the 
letters label blocks of data.}
\label{gridexample}
\end{figure*}
The numbers in that figure indicate the resolution level, and the letters denote blocks. At the base of the 
tree structure is a single block, labeled 1A. Paramesh refines blocks by bisection in each coordinate direction. In 
this one--dimensional example block 1A is refined into two blocks, 2A and 2B. Since each data block contains the 
same number of cells, level 2 has twice the resolution as level 1. Using Paramesh terminology, block 1A is the 
``parent'' of  blocks 2A and 2B, and blocks 2A and 2B are the ``children'' of block 1A . In Fig.~1 Paramesh has 
also refined blocks 2A 
and 2B to create blocks 3A, 3B, 3C and 3D. Further refinements yield the non--uniform grid 
shown in the figure. Paramesh always creates grid structures in which adjacent blocks' refinement levels 
differ by no more than one. 

Our first task is to decide how to carry out a basic multigrid V-cycle on a non--uniform grid structure. 
There are two natural approaches. The first approach is the Fast Adaptive Composite Grid Method (FAC) 
developed by McCormick\cite{McCormickFAC:1986}.  In this approach the relaxation sweeps extend 
across the entire computational domain, and one defines the succession of multigrid levels by restricting the highest 
resolution subgrid to  the next lower resolution. As an example based on Fig.~1, we let the top multigrid 
level consist of blocks 4A-5A-5B-5C-5D-4D-3C-4E-4F. After carrying out a series of relaxation sweeps on this 
non--uniform grid, we step down the multigrid V-cycle by restricting the data in blocks 5A-5B-5C-5D to resolution 
level 4. Thus, the next multigrid level is defined by blocks 4A-4B-4C-4D-3C-4E-4F. After relaxing on this grid, 
we restrict the resolution level 4 blocks to resolution level 3. This defines the next multigrid level as 3A-3B-3C-3D.
We can continue in this fashion 
to define a complete hierarchy of multigrid levels, each covering the entire computational domain. 

The second approach, the one we use for AMRMG, is to define the grids in the multigrid hierarchy to 
coincide with the different resolution levels. This is the original multi--level adaptive technique 
(MLAT) proposed by Brandt \cite{Brandt77,BaiBrandt:1987}.
As an example based on Fig.~1, the top multigrid level consists 
of all blocks at resolution level 5, namely, 5A-5B-5C-5D. After carrying out a series of relaxation sweeps 
on the level 5 blocks, we restrict that data to level 4. Then the next multigrid level consists of blocks 
4A-4B-4C-4D-4E-4F. After relaxing on these blocks we restrict the resolution level 4 data to resolution level 
3. Then the next multigrid level consists of the level 3 blocks 3A-3B-3C-3D. We continue in this fashion 
to define a complete multigrid hierarchy. 

We have built and tested a one--dimensional multigrid code based on the FAC approach. That code works 
quite well. However, the MLAT approach appeared 
to us to be more straightforward to implement in a three--dimensional code based on Paramesh. For this reason AMRMG 
defines the levels in the multigrid hierarchy by resolution.
Apart from the issue of implementation, the MLAT approach has an advantage in solving problems in which only a small 
region of the computational domain requires high resolution.  With the FAC approach, in which relaxation always 
extends across the entire computational domain, a lot of unnecessary computational effort can be expended 
on relaxation in the low resolution regions. On the other hand, the FAC approach has the advantage over MLAT
in maintaining a tighter coupling between regions of different resolutions. For example, for the grid shown in Fig.~1, 
the data in blocks 4A and 4D effectively provide boundary conditions for relaxation in blocks 5A through 5D. With FAC,
that boundary information is updated between every relaxation sweep across the top multigrid level 
(4A-5A-5B-5C-5D-4D-3C-4E-4F). With MLAT, in which the top multigrid level consists of blocks 5A-5B-5C-5D, 
the boundary information is only updated once each V-cycle. 

\section{Guard Cell Filling}\label{sec3}
AMRMG uses cell centered data. When we apply the relaxation formula $f_i(\phi^{\rm new}) = g_i(\phi^{\rm old})$
to a cell adjacent to a block face, the finite difference stencil extends beyond the block. 
Paramesh uses layers of guard cells surrounding each block  to hold data from beyond the 
block boundaries. 
In Fig.~2 we show a portion of a one--dimensional grid 
with a fine grid block on the left and a coarse grid block on the right. 
\begin{figure}[htb]
\includegraphics{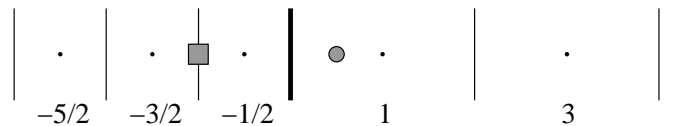}
\caption{A portion of a one--dimensional grid, showing three cells from a fine grid block on the left 
and two cells from a coarse grid block on the right.}
\label{oneDgrid}
\end{figure}
The grid points are labeled by their distance  from the 
block interface in units of the fine grid cell size $\Delta x$. Thus, the fine grid points are 
$-1/2$, $-3/2$, {\it etc}. and the coarse grid points are  $1$, $3$, {\it etc}. The gray circle at 
location $+1/2$ is a guard cell for the fine grid block, and the gray square at location $-1$ is a guard 
cell for the coarse grid block. 
Guard cell values are obtained by interpolation from surrounding interior data points. We want to 
consider how errors in guard cell filling affect the accuracy of the solution. 

Consider the simple example of the Poisson equation in one dimension, $\partial^2\phi/\partial x^2 = \rho$. 
For the moment let us consider a uniform numerical grid with grid spacing $\Delta x$. With standard second 
order centered differencing, the discrete Poisson equation is 
\begin{equation}\label{eqn:Poisson}
  \frac{\phi_{i+1} - 2\phi_i + \phi_{i-1}}{\Delta x^2} + {\cal O}(\Delta x^2) = \rho_i \ .
\end{equation}
where $i$ labels the grid points. 
The term ${\cal O}(\Delta x^2)$ is the truncation error obtained from  discretization 
of the second derivative. We can rewrite Eq.~(\ref{eqn:Poisson}) in a form appropriate for relaxation as
\begin{equation}\label{eqn:relaxPoisson}
  \phi^{\rm new}_i = \frac{1}{2}(\phi^{\rm old}_{i+1} + \phi^{\rm old}_{i-1}) 
  - \frac{1}{2} \Delta x^2 \rho_i + {\cal O}(\Delta x^4) \ .
\end{equation}
When we apply this relaxation formula, the  truncation 
error  dictates that  the numerical solution will have errors of order $\Delta x^2$. 
That is, the numerical solution will be second order accurate. 

For the non--uniform grid of Fig.~2, the discrete equation (\ref{eqn:Poisson}) and the relaxation 
formula (\ref{eqn:relaxPoisson}) apply as shown
in the fine grid region, where $i = -1/2$, $-3/2$, {\it etc}. 
For relaxation  at the grid point $i = -1/2$, we need the guard cell value $\phi_{1/2}$.  We want the 
guard cell value to be sufficiently accurate that it does not spoil the 
second order convergence of the solution.
It is clear from these equations that errors 
of order $\Delta x^4$ in the value of $\phi_{1/2}$ can be absorbed into the truncation 
error already present. 
Thus, we expect the numerical solution to be second order convergent if the guard cells are filled to 
fourth (or higher) order accuracy. 

As far as we know, fourth order guard cell filling is {\it sufficient} to produce a second order accurate 
solution for second order partial differential equations  discretized on a non--uniform grid 
with standard second order differencing. 
Fourth order guard cell filling, however, is not a {\it necessary} condition. 
There is a ``rule of thumb'' in the computational mathematics community that can be summarized as 
follows \cite{BaiBrandt:1987,Henshaw:1990,Martin:1996}: Errors of 
order $\Delta x^p$ that occur on a subspace of dimension $m$ in a space of dimension $n$
will often contribute to the solution like errors of order $\Delta x^{p+n-m}$ from the bulk. Thus, we anticipate that 
errors of order $\Delta x^3$ in guard cell filling, which occur on the two--dimensional block boundaries in the 
three--dimensional space,  will contribute like errors of order $\Delta x^4$ from the bulk
and will not spoil the second order convergence of our code. 
For the problems that we have studied this is indeed the case. 

\begin{figure*}[htb]
\centerline{\includegraphics{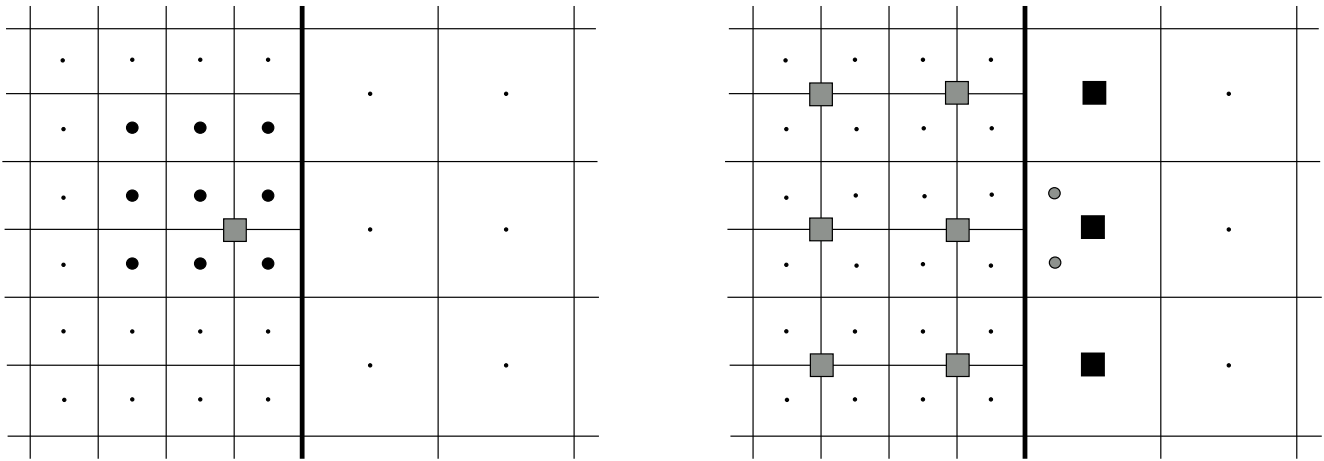}}
\caption{The picture on the left shows the first step in guard cell filling, in which one of the 
parent grid cells (gray square) is filled using quadratic interpolation 
across nine interior fine grid cells (black circles). 
The other parent grid cells are filled using corresponding stencils of 
nine interior fine grid cells. The picture on the right shows the second step in which 
two fine grid guard cells (gray circles) are filled using 
quadratic interpolation across nine parent grid values (squares). 
These parent grid values include one layer of guard cells (black squares) obtained from 
the parent's neighbor on the right side of the interface, and two layers of interior cells (gray squares).}
\label{GCFfigure}
\end{figure*}

The guard cell filling scheme that we use was written by Kevin Olson as part of the standard Paramesh package. 
The process of filling the guard cells of a fine grid block that is adjacent to a coarse grid block 
proceeds in two steps. The first step is a restriction operation in which cells from the interior of the 
fine grid block 
are used to fill the interior cells of the underlying ``parent'' block. 
The restriction operation is depicted for the case of two spatial dimensions 
in the left panel of Fig.~\ref{GCFfigure}.
The restriction proceeds as a succession of one--dimensional quadratic 
interpolations, and is accurate to third order in the grid spacing. 
Note that the fine grid stencil used for this step 
(nine black circles in the figure) cannot be centered on the parent cell
(gray square).  In each dimension the stencil includes two fine 
grid cells on one side of the parent cell and one fine grid cell on the other. 
The stencil is always positioned so that its center is shifted toward 
the center of the block (assumed in the figure to be toward the upper left).  
This ensures that only interior fine grid points, and no fine grid guard cells,
are used in this first step.

For the second step, the fine grid guard cells are filled by prolongation 
from the parent grid. Before the prolongation, the parent block gets 
its own guard cells (black squares in the right panel of Fig.~\ref{GCFfigure}) 
from the neighboring block at the same refinement level.
The stencil used in the prolongation operation is shown in the right panel 
of Fig.~\ref{GCFfigure}.
The prolongation operation proceeds as a succession of one--dimensional 
quadratic interpolations, and is third order accurate. 
In this case, the parent grid stencil includes a layer of guard cells 
(black squares), 
as well as its own interior grid points (gray squares).
At the end of this second step the fine grid guard cells are filled 
to third order accuracy. 

When Paramesh fills the guard cells of a parent block, it also fills the guard cells of the 
parent's neighbor at the same refinement level. In Fig.~3, the parent's neighbor is the coarse 
grid block on the right side of the refinement boundary. Since we are using the second approach 
to multigrid outlined in Sec.~\ref{sec2}, we do not relax in the parent's neighbor block until we step 
down the multigrid hierarchy. Thus, the guard cell values assigned to a parent's neighbor by the 
Paramesh guard cell filling routine are not used by AMRMG.

\section{The FAS algorithm}\label{sec4}
AMRMG uses the Full Approximation Storage (FAS) multigrid algorithm \cite{Brandt77,NumericalRecipes}. 
In this section we provide a brief overview 
of FAS, and discuss the order of restriction and prolongation operators used for stepping down and up
the multigrid hierarchy. 

We are interested in the nonlinear elliptic differential equation $E(\cdot) = \rho$, where $E$ is a 
(possibly) nonlinear elliptic operator. To avoid confusion with notation for exact and approximate 
solutions, we use a centered ``dot'' as a placeholder for the unknown function.
Now consider a simple multigrid hierarchy 
consisting of two grids, a fine grid at level 2 and a coarse grid at level 1. The differential equation 
$E(\cdot) = \rho$,  discretized on the finest level, becomes $E_2(\cdot) = \rho_2$. Here, the subscripts 
denote the multigrid level. What we seek is a solution of the difference equations $E_2(\cdot) = \rho_2$. 

The basic V-cycle for the  FAS algorithm consists of the following steps. 

\noindent {$\underline{\hbox{Step 1}}$}. Guess a trial solution $\tilde\phi_2$ 
(for example, $\tilde\phi_2 = 0$) and carry out some number of 
relaxation sweeps on the equation $E_2(\cdot) = \rho_2$ to obtain an approximate solution $\phi_2$. 

\noindent {$\underline{\hbox{Step 2}}$}. Construct the coarse grid source  
\begin{equation} \label{eqn:step2}
  \rho_1 = {\cal R}(\rho_2 - E_2(\phi_2)) + E_1({\cal R}\phi_2) \ .
\end{equation}
Here ${\cal R}$ denotes the restriction of data from multigrid level 2 to level 1. Also, $E_1$ denotes 
the discretization of the elliptic operator $E$ on the coarse grid 1. Loosely speaking, the term 
$-{\cal R}(E_2(\phi_2))$  removes the 
predominantly short wavelength part of the source that the fine grid has already captured in Step 1. 
The term  $E_1({\cal R}\phi_2)$ returns the long wavelength part of the source that was removed by 
subtracting ${\cal R}(E_2(\phi_2))$.

\noindent {$\underline{\hbox{Step 3}}$}. Start with the trial solution $\tilde\phi_1 = {\cal R}\phi_2$ and carry out 
some number of relaxation sweeps on the equation $E_1(\cdot) = \rho_1$ to obtain an 
approximate solution $\phi_1$. Alternatively, if possible, solve $E_1(\cdot) = \rho_1$ exactly for $\phi_1$. 

\noindent {$\underline{\hbox{Step 4}}$}. Construct the trial solution
\begin{equation} \label{eqn:step4}
  \tilde\phi_2 = \phi_2 + {\cal P}(-{\cal R}\phi_2 + \phi_1) \ .
\end{equation}
Here ${\cal P}$ denotes the prolongation of data from multigrid level 1 to level 2. Loosely speaking, 
the term ${\cal P}(-{\cal R}\phi_2 + \phi_1)$ removes the long wavelength part of $\phi_2$ and replaces it with 
$\phi_1$, which contains primarily long wavelength information due to the construction of $\rho_1$. 

\noindent {$\underline{\hbox{Step 5}}$}. Start with the trial solution $\tilde\phi_2$ from Step 4 and 
carry out some number of relaxation sweeps on the equation $E_2(\cdot) = \rho_2$ to obtain an improved 
approximate solution $\phi_2$. 

For successive V-cycles the approximate solution $\phi_2$ from Step 5 is used as the  trial solution 
$\tilde\phi_2$ in Step 1. The FAS V-cycle can be generalized in an obvious way to any number of 
multigrid levels. 

At the bottom of each V-cycle Step 3 instructs us to find an exact or approximate solution of the equation 
$E_1(\cdot) = \rho_1$. The subscript `$1$' denotes the coarsest level in the multigrid hierarchy. It is 
important for the performance of any multigrid code to solve this equation accurately. 
Solving the level $1$ equation can be a potential bottleneck 
for our code because Paramesh places the data for each block on a single processor. When the 
algorithm is at the bottom of a V-cycle, only a single processor is active.

The simplest 
strategy for solving the level $1$ equation $E_1(\cdot) = \rho_1$ is to carry out  relaxation 
sweeps, just as we do for the higher multigrid levels. Typically our coarsest multigrid level is a single 
data block with $8^3$ interior 
grid points. We find that with Robin boundary conditions it typically takes about one hundred relaxation sweeps 
to solve the level $1$ equation to sufficiently high accuracy.  
With fewer sweeps at this level the code can require more V-cycles to solve the elliptic problem.

The concern is that the solution of the level $1$ equation, requiring $\sim 100$ sweeps with only one 
processor active, can dominate the run time for the code. However, it turns out that the run time for 
AMRMG is dominated by communications calls made in Paramesh. 
Currently AMRMG uses version 3.0 of Paramesh, which is one of 
the first versions of Paramesh to run under MPI. More recent versions are better optimized, but we have 
not switched to the latest version of Paramesh because of the special modifications that AMRMG requires. 

One way that we have improved the performance of our code is to bypass the communications calls made by 
Paramesh when solving the level $1$ equation. The data for the level $1$ equation always resides on 
block $\#1$ on processor $\#1$, so no communication among processors is needed. We have bypassed the 
Paramesh guard cell filling routine by writing a routine that directly fills the guard cells of this 
block using the outer boundary conditions. 

We are primarily interested in solving elliptic problems that are semi--linear, 
that is, problems in which the second order
derivative terms are linear in the unknown field. For these problems the equation to be solved at the bottom of 
each V-cycle takes the form $\triangle_1 \phi_1 = \rho_1$, where $\triangle_1$ is the Laplacian operator 
(not necessarily on flat space with Cartesian coordinates) discretized on multigrid level $1$. In these 
cases we have an alternative to relaxation, namely,  direct 
matrix inversion:  $\phi_1 = (\triangle_1)^{-1}\rho_1$. 
We have implemented matrix inversion for level $1$ using the  direct Gaussian elimination routine from the 
LAPACK libraries \cite{lapack_website}. With Robin boundary conditions, we must solve the 
level $1$ equation for $\phi$ values in the guard 
cells as well as interior cells. With $8^3$ interior grid points and one layer of guard cells, we have $10^3$ values 
to determine at the bottom of each V-cycle. Therefore the matrix to be inverted has dimensions $1000\times 1000$. 
Our tests show that it takes longer (by a factor of $\sim 10$) to solve the level $1$ equation by direct 
matrix inversion than by 
relaxation, assuming the Paramesh communications calls have been bypassed. However, in either case the time 
required to solve the level $1$ equation is a small fraction of the overall runtime for the code. Thus, we 
prefer to use the direct matrix inversion whenever possible 
because, with matrix inversion, the accuracy of the level $1$ solution is insured. 

\section{Restriction and Prolongation}\label{sec5}
As described in Sec.~\ref{sec3}, guard cells are filled with a combination of restriction and prolongation operations. 
The operators used for guard cell filling are third order accurate, and we  denote these by 
${}^{(3)}{\cal R}$ for restriction and ${}^{(3)}{\cal P}$ for prolongation. What order restriction and 
prolongation operators do we use for stepping down and up the multigrid hierarchy? The answer is that 
we use a combination of second,  third, and fourth order operators. 

The restriction operators in Paramesh are always defined in such a way that only interior cells from the 
child blocks are used to fill the interior cells of a parent block. The fine grid stencil is positioned to keep the coarse 
grid point as close as possible to the center of the stencil. For the case of second order restriction, ${}^{(2)}{\cal R}$, 
the coarse grid point lies at the center of the stencil and gets its value from a succession of linear interpolations in 
each dimension.  The case of third order restriction, ${}^{(3)}{\cal R}$, is depicted in the left panel of 
Fig.~\ref{GCFfigure} and is described in Sec.~\ref{sec3}.

The second, third, and fourth order prolongation operators in Paramesh use a succession of (respectively) linear, 
quadratic, and cubic interpolations in each dimension to fill the fine grid cells. 
The prolongation operators use both interior cells and guard cells from a parent block to fill both interior 
and guard cells of child blocks. The right panel of Fig.~\ref{GCFfigure} shows the stencil used 
by the third order prolongation operator, ${}^{(3)}{\cal P}$, to fill fine grid guard cells on the right side of a fine 
grid block. This same stencil is used to 
fill the first layer of interior cells (the layer of interior fine grid cells adjacent to the block boundary). 
For the second and third layer of interior fine grid cells, the stencil is shifted to the left by one coarse grid point. 
This pattern of stencil shifting continues across the right half of the fine grid block until the midpoint of the block
is reached. The stencils used for the left half of the fine grid block are the mirror images of those used for the 
right half.  

The version of Paramesh currently used by AMRMG allows for second and third order restriction, and (in principle) arbitrary 
order prolongation. We have carried out many numerical tests to help us choose among different combinations of 
restriction and prolongation operators for stepping down and up the V-cycles. For all of these tests we used third 
order restriction and third order prolongation to fill the fine grid guard cells, as described in Sec.~\ref{sec3}. 
In our tests we did not consider prolongation orders higher than four.  

As we have presented it, the FAS algorithm uses two 
restriction operators in Step 2, one restriction operator in Step 3, 
and one restriction operator in Step 4. It uses one prolongation operator in Step 4. One could consider distributing 
the (assumed linear) restriction operator through the 
first term in Eq.~(\ref{eqn:step2}) and treating the operators independently. Likewise, one could 
consider distributing prolongation operator through the second term in Eq.~(\ref{eqn:step4}) and treating the
operators independently. 
We have not considered the consequences of splitting these terms. Moreover, AMRMG is written such that the 
calculation ${\cal R}\phi_2$ from Step 2 is used as the trial solution  $\tilde\phi_1 = {\cal R}\phi_2$ for Step 3. 
Thus, we have not tested the consequences of treating these restriction operators independently. Note that the restriction of 
the fine grid solution, ${\cal R}\phi_2$, appears in Step 4 as well as Steps 2 and 3. Our tests show that the 
order of the restriction operator in Step 4 must agree with the order used for ${\cal R}\phi_2$ in Steps 2 and 3. 
If not, the algorithm will often fail to converge in the sense that the residual (defined below) will not decrease 
with successive V-cycles. 

The options that remain for restriction and prolongation operators can be  expressed by 
rewriting Eqs.~(\ref{eqn:step2}) and
(\ref{eqn:step4}):
\begin{subequations}\label{eqn:neweqns}
\begin{eqnarray}
   \rho_1 & = & {}^{(b)}{\cal R}(\rho_2 - E_2(\phi_2)) + E_1({}^{(a)}{\cal R}\phi_2) \ , \label{eqn:neweqnrho}\\
   \tilde\phi_2 & = & \phi_2 + {}^{(c)}{\cal P}(-{}^{(a)}{\cal R}\phi_2 + \phi_1) \label{eqn:neweqnphi}\ .
\end{eqnarray}
\end{subequations}
The letters  $a$, $b$, and $c$  represent the orders of restriction and prolongation operators 
that appear in stepping down and up the V-cycles.

We want to find values for $a$, $b$, and $c$ that will give the best performance. In judging the performance 
of our code we are looking to see how quickly the residual decreases with successive V-cycles for a fixed non--uniform
mesh. The residual at each point in  the computational domain is defined by 
$res = \rho_n - E_n(\phi_n)$, where $n$ is the 
highest refinement level at that point and $\phi_n$ is the approximate solution. The norm of the residual is 
computed as 
\begin{equation} \label{eqn:normres}
   \langle res \rangle = \sqrt{\frac{1}{N}\left(\sum res^2\right)} \ .
\end{equation}
The sum extends over the grid points that cover the computational domain at the highest refinement level. (In 
Fig.~(2) these would be the interior points of blocks 4A-5A-5B-5C-5D-4D-3C-4E-4F.) The number $N$ 
is the total number of such grid points. The norm $\langle res\rangle$ defined above  is similar to the usual $L_2$ norm, but 
lacks a factor of the cell volume in the ``measure'' of the sum. That is, the usual $L_2$ norm would be written 
as $\sqrt{ (\sum v\, res^2)/V}$ where $v$ is the volume of each grid cell and $V$ is the total volume of the computational 
domain. By omitting the factors of cell volume, the norm $\langle res \rangle$   gives equal weighting to the 
residuals in each grid cell, regardless of resolution. 

To be specific, we will quote the results for the simple test problem $\triangle\phi = \rho$, where $\triangle$ is the 
flat space Laplacian in Cartesian coordinates. We use the source $\rho = (6 - 9r^3)\exp(-r^3) $ with 
$r   = \sqrt{x^2 + y^2 + z^2}$. At the boundaries we use the Robin condition $\frac{\partial}{\partial r}
[r(\phi - 1)] = 0$. The analytic
solution to this problem is $\phi = 1 + (1 - \exp(-r^3))/r$. The numerical solution is computed with a 
fixed three--level ``box--in--box'' grid structure. The highest resolution region has 
cell size $\Delta x = 0.125$ and covers a cubical domain with $x$, $y$, and $z$ ranging from $-2$ to $2$.
The medium resolution region has cell size $\Delta x = 0.25$. It covers the domain from $-4$ to $4$ that is 
exterior to the high resolution region. The low resolution region has cell size 
$\Delta x = 0.5$. It covers the domain from $-8$ to $8$ that is exterior to the medium resolution region. 
It has been our experience that the generic 
behavior of AMRMG is fairly well represented by this simple test case. 
 
Our first observation is that with $a = 2$ the residual gets ``stuck'' after a few V-cycles. The norm $\langle res\rangle$ 
drops to about $10^{-4}$, but no further. This happens regardless of the values chosen for $b$ and $c$. In the limit of high 
resolution the truncation error is less than the residual, and the code fails to show second order convergence of 
the solution. Thus, we can eliminate the cases in which $a=2$ and focus on $a=3$. 

For $a=3$ the norm of the residual decreases with successive V-cycles to values well below the truncation error. 
Table \ref{table1} shows the average change in the common logarithm of $\langle res\rangle$ for each V-cycle, as a function of 
the number of 
relaxation sweeps at each multigrid level. (This excludes the first multigrid level, at the bottom of each V-cycle, 
where we compute the exact solution using matrix inversion.) 
\begin{table} [htb]
\begin{tabular}{|c||c|c|c|c|c|c|}\hline
  number  & \multicolumn{6}{|c|}{order of restriction and }\\ 
  of  & \multicolumn{6}{|c|}{prolongation operators ($a$,$b$,$c$)}\\ \cline{2-7}
  sweeps & (3,2,2) & (3,2,3) & (3,2,4) & (3,3,2) & (3,3,3) & (3,3,4) \\ \hline
  1 & $-0.00$ & $-0.11$ & $-0.16$ & $-0.11$ & $-0.13$ & $-0.33$ \\ \hline
  2 & $-0.32$ & $-0.42$ & $-0.50$ & $-0.33$ & $-0.46$ & $-0.56$ \\ \hline
  3 & $-0.56$ & $-0.75$ & $-0.77$ & $-0.43$ & $-0.56$ & $-0.69$ \\ \hline
  4 & $-0.77$ & $-0.88$ & $-1.13$ & $-0.50$ & $-0.63$ & $-0.74$ \\ \hline
  5 & $-0.85$ & $-1.08$ & $-1.21$ & $-0.56$ & $-0.68$ & $-0.79$ \\ \hline
  6 & $-0.87$ & $-1.15$ & $-1.23$ & $-0.62$ & $-0.74$ & $-0.83$ \\ \hline
\end{tabular}
\caption{Average change in $\log(\langle res\rangle)$ per V-cycle.} 
\label{table1}
\end{table}
The best performance is obtained with $(a,b,c) = (3,2,4)$. Note that the norm of the residual becomes
insensitive to the number of relaxation sweeps as the number of sweeps increases beyond four or five.  This is because, 
as observed in Sec.~\ref{sec2}, the higher multigrid levels that have lower resolution neighbors can only receive 
updated boundary information once each V-cycle. It does not help to continue relaxation sweeps when the 
boundary information is ``old'' and needs to be updated. We typically use four relaxation sweeps, with 
red--black Gauss--Seidel ordering \cite{NumericalRecipes}.

The results of our testing lead to the following formulas for Steps 2 and 4 of the FAS algorithm:
\begin{subequations}\label{eqn:finaleqns}
\begin{eqnarray}
   \rho_1 & = & {}^{(2)}{\cal R}(\rho_2 - E_2(\phi_2)) + E_1({}^{(3)}{\cal R}\phi_2) \ , \label{eqn:finaleqnrho}\\
   \tilde\phi_2 & = & \phi_2 + {}^{(4)}{\cal P}(-{}^{(3)}{\cal R}\phi_2 + \phi_1) \label{eqn:finaleqnphi}\ .
\end{eqnarray}
\end{subequations}
In Step 3 we use the trial solution $\tilde\phi_1 = {}^{(3)}{\cal R} \phi_2$. Recall that we have not tested
the algorithm with order of restriction greater than $3$, or with order of prolongation greater than $4$. 

Conventional wisdom for determining the orders of restriction and prolongation used for
multigrid transfer operations is that the following should be satisfied:
\begin{equation}\label{eqn:orders}
{\cal O}_{\cal R} + {\cal O}_{\cal P} >  {\cal O}_{\cal D} \ .
\end{equation}
Here, ${\cal O}_{\cal R}$, ${\cal O}_{\cal P}$, and ${\cal O}_{\cal D}$ are the orders of restriction, 
prolongation, and the differential operator, respectively \cite{MultigridTextBook}. 
For a uniform grid,  AMRMG acts as a typical FAS multigrid solver and we achieve acceptable convergence rates 
as long as the transfer operators satisfy Eq.~(\ref{eqn:orders}).  With a nonuniform grid, the restriction 
operator denoted ${}^{(a)}{\cal R}$ in Eqs.~(\ref{eqn:neweqns}) must be third or higher order, at 
least in the vicinity of mesh refinement boundaries, for the code to achieve both second order accuracy 
and optimal convergence rates. This is not surprising since, with the MLAT approach, data that is 
restricted from a high resolution 
region (for example, data restricted from block 5D to block 4C in Fig.~1) can serve as boundary data for 
relaxation in a coarse grid region (block 4D in Fig.~1). With restriction order less than $3$, such boundary 
data yield truncation errors greater than ${\cal O}(\Delta x)$ when they appear in a discrete
second derivative. One of our goals for AMRMG is to achieve second order accuracy and good convergence rates with 
minimal modification of the existing Paramesh framework. For this reason we have not 
explored the possibility of using modified finite difference stencils or 
modified transfer operators in the vicinity of mesh refinement boundaries. 

\section{Truncation Error and Grid Control}\label{sec6}
AMRMG adapts the grid structure to the  problem at hand in an attempt to keep the local truncation error under control. 
The local truncation error is defined across the computational domain on the grid that consists of the 
highest resolution blocks. In Fig.~(2) these would be blocks 4A-5A-5B-5C-5D-4D-3C-4E-4F. Let us refer to 
this non--uniform grid as grid $h$. Then the local truncation error is defined by \cite{Brandt77}
\begin{equation}\label{eqn:truncerr}
   \tau_h = E_h(\phi|_h) - \left.\left(E(\phi)\right)\right|_h \ ,
\end{equation}
where $\phi$ is the exact solution of the continuum equation $E(\phi) = \rho$, and $\phi|_h$ is the projection 
of $\phi$ onto grid $h$. The discretization of the operator $E$ on grid $h$ is denoted by $E_h$. 
In a similar manner we define the local truncation error $\tau_H$ on a grid $H$ that is 
constructed from the parents of grid $h$ blocks. Grid $H$ covers the computational domain with half the 
resolution of grid $h$. The difference between the truncation errors on grids $H$ and $h$  is 
\begin{eqnarray} \label{eqn:taudiff}
   \tau_H - {\cal R}\tau_h & = & E_H(\phi|_H) - (E(\phi))|_H \nonumber\\
  & &\quad - {\cal R}E_h(\phi|_h) + {\cal R}(E(\phi))|_h \ ,
\end{eqnarray}
where ${\cal R}$ is a linear operator that restricts data from $h$ to $H$. Let us assume that 
${\cal R}$ is third  
order accurate in the grid spacing. Then the second and fourth terms in Eq.~(\ref{eqn:taudiff}) 
cancel  to third order and, to this same order of accuracy,  we find 
$\tau_H - {\cal R}\tau_h \approx E_H(\phi|_H) - {\cal R}E_h(\phi|_h)$.

The relative local truncation error is defined on grid $H$ by \cite{Brandt77}
\begin{equation}\label{eqn:reltruncerr}
   \tau^H_h = E_H({\cal R}\tilde\phi_h) - {\cal R} ( E_h (\tilde\phi_h)) \ ,
\end{equation}
where $\tilde\phi_h$ is the approximate numerical solution from grid $h$. Again we assume that the restriction 
operator is accurate to third order. Since the approximate solution 
coincides with the exact solution to leading order, $\tilde\phi_h \approx \phi|_h$, we see from Eqs.~(\ref{eqn:taudiff}) 
and (\ref{eqn:reltruncerr}) that 
to leading order in the grid spacing the relative truncation error is related to the local truncation errors by 
$\tau^H_h \approx \tau_H - {\cal R}\tau_h$.
Since we use second order differencing for our elliptic problems  the truncation errors are proportional to the 
square of the grid spacing. Then $\tau_H \approx 4\,{\cal R}\tau_h$ and the relative truncation error is given by 
$\tau^H_h \approx 3\, {\cal R}\tau_h$. This relation can be prolonged to the finest grid $h$, giving 
${\cal P}\tau^H_h \approx 3\, {\cal P}{\cal R}\tau_h$. The prolongation operator 
${\cal P}$, like the restriction operator ${\cal R}$, is assumed to be third order accurate in the grid spacing. 
Then to third order accuracy ${\cal P}{\cal R}$ 
is the identity operator on $h$, and  we have 
\begin{equation} \label{eqn:truncerrapprox1}
   \tau_h \approx \frac{1}{3} {\cal P}\tau^H_h \ .
\end{equation}
Together, Eqs.~(\ref{eqn:reltruncerr}) and (\ref{eqn:truncerrapprox1}) give 
\begin{equation}\label{eqn:truncerrapprox}
   \tau_h \approx \frac{1}{3} \left( {}^{(3)}{\cal P} E_H({}^{(3)}{\cal R}\tilde\phi_h) - 
         E_h(\tilde\phi_h) \right) \ .
\end{equation}
In AMRMG, we use this approximation of the local truncation error to monitor the errors and 
control the grid structure. Note that since the truncation error $\tau_h$ is proportional to the square of 
the grid spacing, the result (\ref{eqn:truncerrapprox}) is valid to leading order only if third (or higher) order 
restriction and prolongation operators are used.  

\begin{figure} \includegraphics[scale=1]{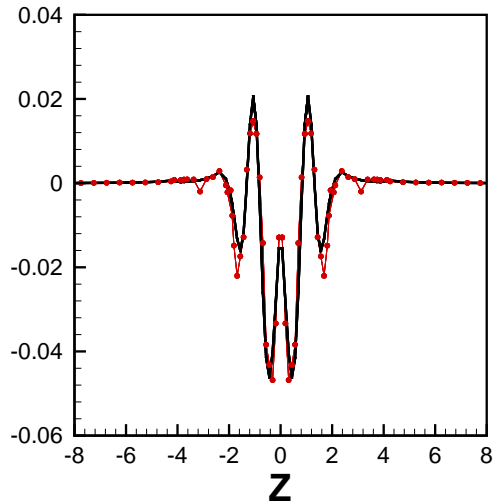}
\caption{Analytic truncation error and calculated truncation error along the $z$--axis.}
\label{rteplots}
\end{figure}

For the test problem described in Section \ref{sec5}, it is straightforward to calculate the analytic truncation error.  Figure \ref{rteplots} shows a comparison of the analytic truncation error with the computed approximation to the truncation error (\ref{eqn:truncerrapprox}) for that test problem.  The analytic truncation error is shown as a thick solid line, while the calculated truncation error is a thin line with filled circles.

In practice we start the solution process by specifying a rather coarse grid structure, sometimes uniform 
sometimes not. The code V-cycles until the norm of the residual across grid $h$ is less than the norm of the 
local truncation error across grid $h$. We write this as  $\langle res \rangle|_h < \langle \tau_h \rangle|_h$, 
where the norm of the 
truncation error is defined in the same way as the norm of the residual, Eq.~(\ref{eqn:normres}). The code 
usually requires  two or three V-cycles to meet this criterion. 
We then compute the norm of the truncation error $\langle \tau_h \rangle|_b$  
for each block of grid $h$. Any block whose norm is greater than some threshold value, 
$\langle \tau_h \rangle|_b > \tau_{\rm max}$, is flagged for refinement. Paramesh rebuilds the grid 
structure and redistributes the data across processors. To obtain a trial solution in the newly formed 
blocks we prolong the solution from the parent blocks. The code then carries out V-cycles on 
this new grid structure with its new highest--resolution grid $h$. The entire process repeats until all blocks 
satisfy $\langle \tau_h \rangle|_b \leq \tau_{\rm max}$  and no blocks are 
flagged for refinement. At this point the code continues to V-cycle until two conditions are satisfied: 
(1) the norm of the residual in each block is less than the norm of the truncation error,  
$\langle res \rangle|_b < \langle \tau_h \rangle|_b$; and (2) the norm of the residual across the entire 
grid is less than some threshold value, $\langle res \rangle|_h < res_{\rm max}$. If only the first condition 
is desired, we simply set $res_{\rm max}$ to a very large value. 

We have tested the code using second, third, and 
fourth order prolongation operators for the calculation of trial solutions in newly formed blocks. We find that 
none of these operators is consistently better than the others. The entire adaptive mesh, multigrid algorithm is 
not very sensitive to the order of prolongation used in this step. We 
typically use the fourth order operator ${}^{(4)}{\cal P}$. 

The grid control scheme used by AMRMG  works well. It insures that the truncation error in each block 
across the computational domain is uniformly low, less than $\tau_{\rm max}$, and 
that the errors coming from the residuals in each block are less than the truncation errors. 
The value chosen for $\tau_{\rm max}$ depends on the problem being solved and the desired 
degree of accuracy. 

\section{Code Tests}\label{sec7}
\begin{figure}
\includegraphics[scale=.5]{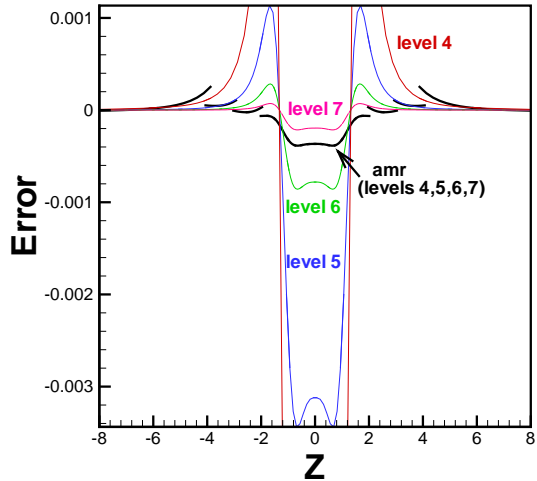}
\caption{Errors along the $z$--axis for AMR grid and uniform grids of various resolutions.}
\label{FMRvsUNI}
\end{figure}

In this section we present code tests to demonstrate second order convergence  
and the computational advantages of AMR.
Figure \ref{FMRvsUNI} shows a comparison of errors for the test case described in Sec.~\ref{sec5}. 
\begin{figure}
\includegraphics{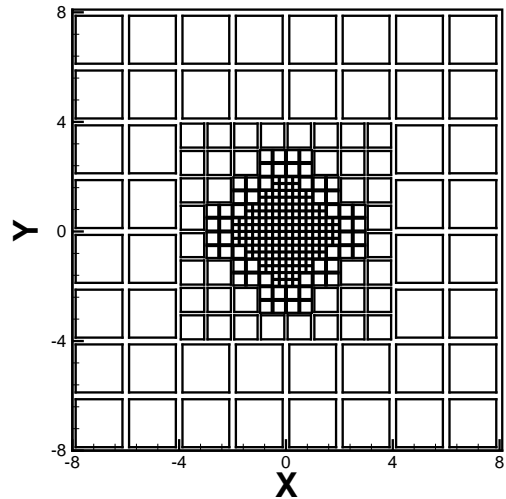}
\caption{AMR grid structure in the $x$-$y$ plane. Each square corresponds to a block of data containing 
$8^3$ computational cells.}
\label{amrgrid}
\end{figure}
The analytical errors on the $z$--axis are plotted for an AMR grid and for uniform grids with resolutions 
at levels $4$, $5$, $6$, and  $7$. (A grid with resolution level  $X$ 
is created by refining a single block  $X-1$ times. See Fig.~1.)
For the AMR run we start with a uniform level $3$ grid and set the refinement criterion for a maximum  
truncation error of  $\tau_{\rm max} = 0.001$ in each block. We also limited the highest resolution to level $7$, so 
the truncation errors in some of the level $7$ blocks reached as high as $0.007$. The 
cell size for resolution level $4$ is $\Delta x = 0.25$, and the cell size for resolution level $7$ is 
$\Delta x = 0.03125$. The grid structure chosen by AMRMG for the AMR run is shown in Fig.~\ref{amrgrid}.

Observe that the errors in each region of the AMR grid are comparable to the errors obtained 
with a uniform grid of the same resolution. For example, the errors in the level $7$ region of the AMR solution 
(the region surrounding the origin) are  slightly larger than the errors obtained from the uniform level $7$ 
solution, and smaller than the errors obtained from the uniform level $6$ solution. We also note that the 
savings in memory with  AMR is profound; the AMR grid solution can be calculated on a single processor, while
the level $7$ uniform grid solution required $64$ processors to handle the memory requirements. 

As a test for second order convergence, we  consider the initial data for a single black hole.
Valid initial data must satisfy the constraint equations of general relativity. For a vacuum spacetime, the
Hamiltonian and momentum constraints are given respectively by
\begin{equation}
R + K^2 - K_{ij}K^{ij} = 0 \ ,
\end{equation}
and
\begin{equation}
D_j(K^{ij} - g^{ij}K) = 0 \ ,
\end{equation}
where $g_{ij}$ is the physical metric and $g^{ij}$ is its inverse. Also, $R$ is the scalar curvature, 
$K_{ij}$ is the extrinsic curvature with trace $K = K^i_i$, and $D_j$ is the covariant derivative 
associated with the spatial metric. These initial value
equations must be rewritten as a well--posed elliptic boundary value problem. The standard techniques for 
rewriting the constraint equations are 
based on the York--Lichnerowicz conformal decomposition \cite{Cook:2000vr}. Following this approach, we 
assume that the physical  metric $g_{ij}$  is conformally related to a background metric $\tilde g_{ij}$,
\begin{equation}
g_{ij} = \psi^4 \tilde g_{ij} \ ,
\end{equation}
where $\psi^4$ is the conformal factor. The physical extrinsic curvature is written as
\begin{equation}
K_{ij} = \psi^{-2} \tilde K_{ij} \ .
\end{equation}
In terms of these conformal variables, the Hamiltonian and momentum constraints read
\begin{subequations}\label{eqn:constraints}
\begin{eqnarray}
  & 8\tilde\nabla^2 \psi - \psi \tilde R + \psi^{-7} \left( \tilde K_{ij} \tilde K^{ij} 
       - \tilde K^2\right)  = 0 \ ,\label{eqn:Hconstraint}\\
  & \tilde\nabla_i(\tilde K^{ij} - {\tilde g}^{ij}\tilde K )  
       = -4\psi^{-1}\tilde K \tilde\nabla^i \psi \ ,\label{eqn:Momconstraint}
\end{eqnarray}
\end{subequations}
where $\tilde\nabla_i$ and $\tilde R$ are the covariant derivative and scalar curvature associated 
with the background metric $\tilde g_{ij}$. 

The ``puncture method'' \cite{Brandt:1997tf} is a way of specifying black hole initial data on 
${\hbox{$I$\kern-3.8pt $R$}}^3$.  The background metric is chosen to be flat.
The momentum constraint (\ref{eqn:Momconstraint}) is solved analytically  by \cite{Bowen:1980yu}
\begin{eqnarray}\label{eqn:bowenyork}
\tilde K^{ij} = \frac{3}{2r^2}(P^in^j+P^jn^i-(\tilde g^{ij}-n^in^j)P^kn_k)
\end{eqnarray}
where $P^i$ is the momentum of the black hole and $n^i$ is the radial normal vector (in the flat 
background with Cartesian coordinates).  Note that $\tilde K^{ij}$ is traceless, $\tilde K = 0$. This  
expression (\ref{eqn:bowenyork}) for $\tilde K^{ij}$  can be generalized to include an arbitrary
number of black holes with spin and momentum, but for simplicity we will use a single black hole with 
no spin for our test case. To complete the specification of initial data, we must solve 
the Hamiltonian constraint (\ref{eqn:Hconstraint})  for the conformal factor $\psi$. With the puncture 
method, the solution $\psi$ is split into a known singular term and a nonsingular term $u$:
\begin{equation}
 \psi = u + \frac{m}{2|{\vec r}|} \ .
\end{equation}
Here, $m$ is the ``bare mass'' of the black hole and $|{\vec r}|$ is the coordinate distance from the origin. 
With the puncture method splitting of the conformal factor, the Hamiltonian constraint becomes
\begin{equation}
\label{ham}
  {\nabla}^2 u +\beta(1+\alpha u)^{-7} = 0 \ ,
\end{equation}
where               
\begin{equation}
 \beta = \frac{1}{8}\alpha^7 \tilde K^{ij} \tilde K_{ij} \quad {\rm and} \quad \alpha = \frac{2|{\vec r}|}{m} \ .
\end{equation}
Equation (\ref{ham}) is solved for the nonsingular function $u$ on ${\hbox{$I$\kern-3.8pt $R$}}^3$ with 
Robin boundary conditions $\frac{\partial}{\partial r}[r(\phi - 1)] = 0$.

For these tests we use a fixed mesh refinement (FMR) grid with an  ``$X$ plus $3$'' ($Xp3$) structure. 
The terminology $Xp3$ means that the grid is composed of $4$ refinement regions,  with the coarsest 
part of the grid at level $X$ and the finest part of the grid  at level $X+3$. The different 
levels are nested in a ``box--in--box'' fashion. The finest level, with resolution $X+3$, extends 
from $-2$ to $2$ in each coordinate direction. The level with resolution $X+2$ covers the domain 
between $-4$ and $4$, excluding the finest level. The level with resolution $X+1$ covers the 
domain between $-8$ and $8$ excluding the finer levels. The coarsest level, with resolution $X$, 
covers the domain between $-16$ and $16$ excluding the finer levels. 

\begin{figure}
\includegraphics{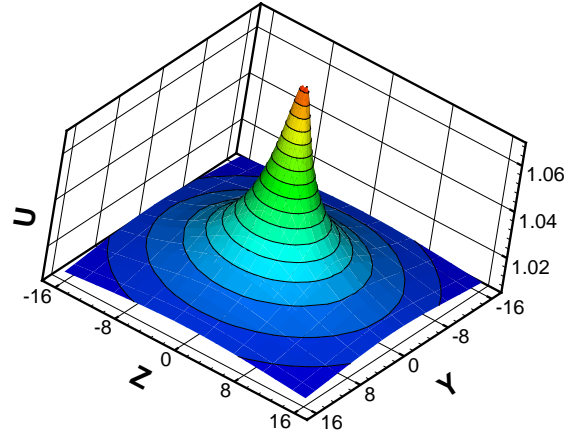}
\caption{Contour plot of the non-singular part of solution $u$.}
\label{Punct}
\end{figure}

For our test case we have chosen $m=1$ and $P^i=(0,0,1)$. We solve Eq.~(\ref{ham}) using the 
series of FMR grids $3p3$, $4p3$, $5p3$, $6p3$, and $7p3$. Each successive FMR grid has
the same boundaries and double the resolution of the previous grid.
Figure \ref{Punct} shows a contour plot of the solution  $u$ in the 
$y$-$z$ plane, obtained with the $6p3$ grid. 

\begin{figure}
\includegraphics[scale=.5]{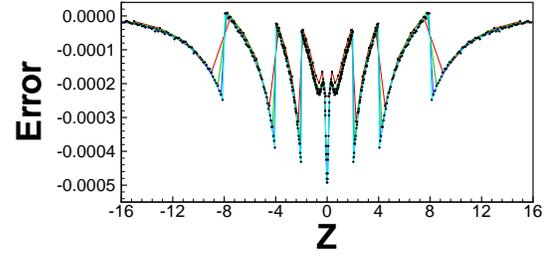}
\caption{Three point convergence test of puncture data.}
\label{PunCon}
\end{figure}

\begin{figure}
\includegraphics[scale=.5]{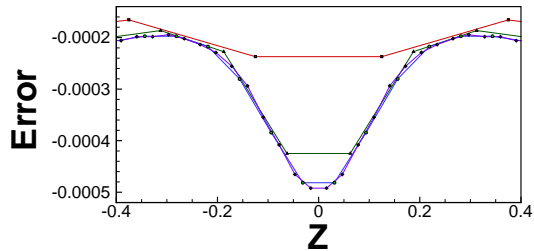}
\caption{Three point convergence test of puncture data close to puncture.}
\label{PunConZoom}
\end{figure}

Figures \ref{PunCon} and \ref{PunConZoom} show the results of a three--point convergence test for 
data along the $z$ axis.  
This data passes through the puncture (at the origin) where the solution  
$u$ and its derivatives are changing most rapidly. 
The three--point convergence test is obtained by plotting the difference between solutions on 
successive FMR grids, multiplied by an appropriate power of $4$. The top (red) curve shown 
in Figs.~\ref{PunCon} and \ref{PunConZoom} is the 
difference between the solution $u$ obtained on the $4p3$ grid and the solution obtained on the $3p3$ 
grid. The next curve  is the difference between solutions on FMR grids $5p3$ and 
$4p3$, multiplied by $4$. The third curve is the difference between solutions on 
FMR grids $6p3$ and $5p3$, multiplied by $16$. Finally, the curve that has the most negative value at the 
origin in Fig.~\ref{PunConZoom} is the difference between solutions on the FMR grids $7p3$ and $6p3$, 
multiplied by $64$. One can see that the curves in Figs.~\ref{PunCon} and \ref{PunConZoom} 
overlay one another  in the limit of high resolution. This shows that the errors in AMRMG are 
second order in the grid spacing. 


\begin{acknowledgments}
We would like to thank the numerical relativity group at NASA Goddard Space Flight Center, and especially Dae-Il Choi, 
for their help and support. We would also like to thank Kevin Olson and Peter MacNeice for 
their help with Paramesh.  This work was supported by NASA Space Sciences Grant ATP02-0043-0056. 
Computations were carried out 
on the North Carolina State University IBM Blade Center Linux Cluster.  
The Paramesh software used in this work was developed at NASA Goddard Space Flight Center 
under the HPCC and ESTO/CT projects. 
\end{acknowledgments}
\bibliography{paper_references}

\begin{thebibliography}{38}
\expandafter\ifx\csname natexlab\endcsname\relax\def\natexlab#1{#1}\fi
\expandafter\ifx\csname bibnamefont\endcsname\relax
  \def\bibnamefont#1{#1}\fi
\expandafter\ifx\csname bibfnamefont\endcsname\relax
  \def\bibfnamefont#1{#1}\fi
\expandafter\ifx\csname citenamefont\endcsname\relax
  \def\citenamefont#1{#1}\fi
\expandafter\ifx\csname url\endcsname\relax
  \def\url#1{\texttt{#1}}\fi
\expandafter\ifx\csname urlprefix\endcsname\relax\def\urlprefix{URL }\fi
\providecommand{\bibinfo}[2]{#2}
\providecommand{\eprint}[2][]{\url{#2}}

\bibitem[{\citenamefont{Choptuik}(1993)}]{Choptuik:1992jv}
\bibinfo{author}{\bibfnamefont{M.~W.} \bibnamefont{Choptuik}},
  \bibinfo{journal}{Phys. Rev. Lett.} \textbf{\bibinfo{volume}{70}},
  \bibinfo{pages}{9} (\bibinfo{year}{1993}).

\bibitem[{\citenamefont{Br{\"u}gmann}(1996)}]{Brugmann:1996kz}
\bibinfo{author}{\bibfnamefont{B.}~\bibnamefont{Br{\"u}gmann}},
  \bibinfo{journal}{Phys. Rev.} \textbf{\bibinfo{volume}{D54}},
  \bibinfo{pages}{7361} (\bibinfo{year}{1996}), \eprint{gr-qc/9608050}.

\bibitem[{\citenamefont{Br{\"u}gmann}(1999)}]{Brugmann:1997uc}
\bibinfo{author}{\bibfnamefont{B.}~\bibnamefont{Br{\"u}gmann}},
  \bibinfo{journal}{Int. J. Mod. Phys.} \textbf{\bibinfo{volume}{D8}},
  \bibinfo{pages}{85} (\bibinfo{year}{1999}), \eprint{gr-qc/9708035}.

\bibitem[{\citenamefont{Papadopoulos et~al.}(1998)\citenamefont{Papadopoulos,
  Seidel, and Wild}}]{Papadopoulos:1998dk}
\bibinfo{author}{\bibfnamefont{P.}~\bibnamefont{Papadopoulos}},
  \bibinfo{author}{\bibfnamefont{E.}~\bibnamefont{Seidel}}, \bibnamefont{and}
  \bibinfo{author}{\bibfnamefont{L.}~\bibnamefont{Wild}},
  \bibinfo{journal}{Phys. Rev.} \textbf{\bibinfo{volume}{D58}},
  \bibinfo{pages}{084002} (\bibinfo{year}{1998}), \eprint{gr-qc/9802069}.

\bibitem[{\citenamefont{Diener et~al.}(2000)\citenamefont{Diener, Jansen,
  Khokhlov, and Novikov}}]{Diener:1999mf}
\bibinfo{author}{\bibfnamefont{P.}~\bibnamefont{Diener}},
  \bibinfo{author}{\bibfnamefont{N.}~\bibnamefont{Jansen}},
  \bibinfo{author}{\bibfnamefont{A.}~\bibnamefont{Khokhlov}}, \bibnamefont{and}
  \bibinfo{author}{\bibfnamefont{I.}~\bibnamefont{Novikov}},
  \bibinfo{journal}{Class. Quant. Grav.} \textbf{\bibinfo{volume}{17}},
  \bibinfo{pages}{435} (\bibinfo{year}{2000}), \eprint{gr-qc/9905079}.

\bibitem[{\citenamefont{New et~al.}(2000)\citenamefont{New, Choi, Centrella,
  MacNeice, Huq, and Olson}}]{New:2000dz}
\bibinfo{author}{\bibfnamefont{K.~C.~B.} \bibnamefont{New}},
  \bibinfo{author}{\bibfnamefont{D.-I.} \bibnamefont{Choi}},
  \bibinfo{author}{\bibfnamefont{J.~M.} \bibnamefont{Centrella}},
  \bibinfo{author}{\bibfnamefont{P.}~\bibnamefont{MacNeice}},
  \bibinfo{author}{\bibfnamefont{M.}~\bibnamefont{Huq}}, \bibnamefont{and}
  \bibinfo{author}{\bibfnamefont{K.}~\bibnamefont{Olson}},
  \bibinfo{journal}{Phys. Rev.} \textbf{\bibinfo{volume}{D62}},
  \bibinfo{pages}{084039} (\bibinfo{year}{2000}), \eprint{gr-qc/0007045}.

\bibitem[{\citenamefont{Choptuik
  et~al.}(2003{\natexlab{a}})\citenamefont{Choptuik, Hirschmann, Liebling, and
  Pretorius}}]{Choptuik:2003ac}
\bibinfo{author}{\bibfnamefont{M.~W.} \bibnamefont{Choptuik}},
  \bibinfo{author}{\bibfnamefont{E.~W.} \bibnamefont{Hirschmann}},
  \bibinfo{author}{\bibfnamefont{S.~L.} \bibnamefont{Liebling}},
  \bibnamefont{and}
  \bibinfo{author}{\bibfnamefont{F.}~\bibnamefont{Pretorius}},
  \bibinfo{journal}{Phys. Rev.} \textbf{\bibinfo{volume}{D68}},
  \bibinfo{pages}{044007} (\bibinfo{year}{2003}{\natexlab{a}}),
  \eprint{gr-qc/0305003}.

\bibitem[{\citenamefont{Choi et~al.}(2004)\citenamefont{Choi, Brown, Imbiriba,
  Centrella, and MacNeice}}]{Choi:2003ba}
\bibinfo{author}{\bibfnamefont{D.-I.} \bibnamefont{Choi}},
  \bibinfo{author}{\bibfnamefont{J.~D.} \bibnamefont{Brown}},
  \bibinfo{author}{\bibfnamefont{B.}~\bibnamefont{Imbiriba}},
  \bibinfo{author}{\bibfnamefont{J.}~\bibnamefont{Centrella}},
  \bibnamefont{and} \bibinfo{author}{\bibfnamefont{P.}~\bibnamefont{MacNeice}},
  \bibinfo{journal}{J. Comput. Phys.} \textbf{\bibinfo{volume}{193}},
  \bibinfo{pages}{398} (\bibinfo{year}{2004}), \eprint{physics/0307036}.

\bibitem[{\citenamefont{Schnetter et~al.}(2004)\citenamefont{Schnetter, Hawley,
  and Hawke}}]{Schnetter:2003rb}
\bibinfo{author}{\bibfnamefont{E.}~\bibnamefont{Schnetter}},
  \bibinfo{author}{\bibfnamefont{S.~H.} \bibnamefont{Hawley}},
  \bibnamefont{and} \bibinfo{author}{\bibfnamefont{I.}~\bibnamefont{Hawke}},
  \bibinfo{journal}{Class. Quant. Grav.} \textbf{\bibinfo{volume}{21}},
  \bibinfo{pages}{1465} (\bibinfo{year}{2004}), \eprint{gr-qc/0310042}.

\bibitem[{\citenamefont{Br{\"u}gmann et~al.}(2004)\citenamefont{Br{\"u}gmann,
  Tichy, and Jansen}}]{Bruegmann:2003aw}
\bibinfo{author}{\bibfnamefont{B.}~\bibnamefont{Br{\"u}gmann}},
  \bibinfo{author}{\bibfnamefont{W.}~\bibnamefont{Tichy}}, \bibnamefont{and}
  \bibinfo{author}{\bibfnamefont{N.}~\bibnamefont{Jansen}},
  \bibinfo{journal}{Phys. Rev. Lett.} \textbf{\bibinfo{volume}{92}},
  \bibinfo{pages}{211101} (\bibinfo{year}{2004}), \eprint{gr-qc/0312112}.

\bibitem[{\citenamefont{Imbiriba et~al.}(2004)\citenamefont{Imbiriba, Baker,
  Choi, Centrella, Fiske, Brown, van Meter, and Olson}}]{Imbiriba:2004tp}
\bibinfo{author}{\bibfnamefont{B.}~\bibnamefont{Imbiriba}},
  \bibinfo{author}{\bibfnamefont{J.}~\bibnamefont{Baker}},
  \bibinfo{author}{\bibfnamefont{D.-I.} \bibnamefont{Choi}},
  \bibinfo{author}{\bibfnamefont{J.}~\bibnamefont{Centrella}},
  \bibinfo{author}{\bibfnamefont{D.~R.} \bibnamefont{Fiske}},
  \bibinfo{author}{\bibfnamefont{J.~D.} \bibnamefont{Brown}},
  \bibinfo{author}{\bibfnamefont{J.~R.} \bibnamefont{van Meter}},
  \bibnamefont{and} \bibinfo{author}{\bibfnamefont{K.}~\bibnamefont{Olson}}
  (\bibinfo{year}{2004}), \eprint{gr-qc/0403048}.

\bibitem[{\citenamefont{Misner et~al.}(1973)\citenamefont{Misner, Thorne, and
  Wheeler}}]{MTW}
\bibinfo{author}{\bibfnamefont{C.}~\bibnamefont{Misner}},
  \bibinfo{author}{\bibfnamefont{K.}~\bibnamefont{Thorne}}, \bibnamefont{and}
  \bibinfo{author}{\bibfnamefont{J.~A.} \bibnamefont{Wheeler}},
  \emph{\bibinfo{title}{Gravitation}} (\bibinfo{publisher}{W.H. Freeman},
  \bibinfo{address}{San Francisco}, \bibinfo{year}{1973}).

\bibitem[{\citenamefont{Cook}(2000)}]{Cook:2000vr}
\bibinfo{author}{\bibfnamefont{G.~B.} \bibnamefont{Cook}},
  \bibinfo{journal}{Living Rev. Rel.} \textbf{\bibinfo{volume}{3}},
  \bibinfo{pages}{5} (\bibinfo{year}{2000}), \eprint{gr-qc/0007085}.

\bibitem[{\citenamefont{Kidder et~al.}(2001)\citenamefont{Kidder, Scheel, and
  Teukolsky}}]{Kidder:2001tz}
\bibinfo{author}{\bibfnamefont{L.~E.} \bibnamefont{Kidder}},
  \bibinfo{author}{\bibfnamefont{M.~A.} \bibnamefont{Scheel}},
  \bibnamefont{and} \bibinfo{author}{\bibfnamefont{S.~A.}
  \bibnamefont{Teukolsky}}, \bibinfo{journal}{Phys. Rev.}
  \textbf{\bibinfo{volume}{D64}}, \bibinfo{pages}{064017}
  (\bibinfo{year}{2001}), \eprint{gr-qc/0105031}.

\bibitem[{\citenamefont{Lindblom and Scheel}(2002)}]{Lindblom:2002et}
\bibinfo{author}{\bibfnamefont{L.}~\bibnamefont{Lindblom}} \bibnamefont{and}
  \bibinfo{author}{\bibfnamefont{M.~A.} \bibnamefont{Scheel}},
  \bibinfo{journal}{Phys. Rev.} \textbf{\bibinfo{volume}{D66}},
  \bibinfo{pages}{084014} (\bibinfo{year}{2002}), \eprint{gr-qc/0206035}.

\bibitem[{\citenamefont{Scheel et~al.}(2002)\citenamefont{Scheel, Kidder,
  Lindblom, Pfeiffer, and Teukolsky}}]{Scheel:2002yj}
\bibinfo{author}{\bibfnamefont{M.~A.} \bibnamefont{Scheel}},
  \bibinfo{author}{\bibfnamefont{L.~E.} \bibnamefont{Kidder}},
  \bibinfo{author}{\bibfnamefont{L.}~\bibnamefont{Lindblom}},
  \bibinfo{author}{\bibfnamefont{H.~P.} \bibnamefont{Pfeiffer}},
  \bibnamefont{and} \bibinfo{author}{\bibfnamefont{S.~A.}
  \bibnamefont{Teukolsky}}, \bibinfo{journal}{Phys. Rev.}
  \textbf{\bibinfo{volume}{D66}}, \bibinfo{pages}{124005}
  (\bibinfo{year}{2002}), \eprint{gr-qc/0209115}.

\bibitem[{\citenamefont{Lindblom et~al.}(2004)}]{Lindblom:2004gd}
\bibinfo{author}{\bibfnamefont{L.}~\bibnamefont{Lindblom}}
  \bibnamefont{et~al.}, \bibinfo{journal}{Phys. Rev.}
  \textbf{\bibinfo{volume}{D69}}, \bibinfo{pages}{124025}
  (\bibinfo{year}{2004}), \eprint{gr-qc/0402027}.

\bibitem[{\citenamefont{Choptuik
  et~al.}(2003{\natexlab{b}})\citenamefont{Choptuik, Hirschmann, Liebling, and
  Pretorius}}]{Choptuik:2003as}
\bibinfo{author}{\bibfnamefont{M.~W.} \bibnamefont{Choptuik}},
  \bibinfo{author}{\bibfnamefont{E.~W.} \bibnamefont{Hirschmann}},
  \bibinfo{author}{\bibfnamefont{S.~L.} \bibnamefont{Liebling}},
  \bibnamefont{and}
  \bibinfo{author}{\bibfnamefont{F.}~\bibnamefont{Pretorius}},
  \bibinfo{journal}{Class. Quant. Grav.} \textbf{\bibinfo{volume}{20}},
  \bibinfo{pages}{1857} (\bibinfo{year}{2003}{\natexlab{b}}),
  \eprint{gr-qc/0301006}.

\bibitem[{\citenamefont{Anderson and Matzner}(2003)}]{Anderson:2003dz}
\bibinfo{author}{\bibfnamefont{M.}~\bibnamefont{Anderson}} \bibnamefont{and}
  \bibinfo{author}{\bibfnamefont{R.~A.} \bibnamefont{Matzner}}
  (\bibinfo{year}{2003}), \eprint{gr-qc/0307055}.

\bibitem[{\citenamefont{Matzner}(2005)}]{Matzner:2004uu}
\bibinfo{author}{\bibfnamefont{R.~A.} \bibnamefont{Matzner}},
  \bibinfo{journal}{Phys. Rev.} \textbf{\bibinfo{volume}{D71}},
  \bibinfo{pages}{024011} (\bibinfo{year}{2005}), \eprint{gr-qc/0408003}.

\bibitem[{\citenamefont{Fedorenko}(1962)}]{Fedorenko:1962}
\bibinfo{author}{\bibfnamefont{R.}~\bibnamefont{Fedorenko}},
  \bibinfo{journal}{USSR Comp. Math. Math. Phys.} \textbf{\bibinfo{volume}{1}},
  \bibinfo{pages}{1092} (\bibinfo{year}{1962}).

\bibitem[{\citenamefont{Fedorenko}(1964)}]{Fedorenko:1964}
\bibinfo{author}{\bibfnamefont{R.}~\bibnamefont{Fedorenko}},
  \bibinfo{journal}{USSR Comp. Math. Math. Phys.} \textbf{\bibinfo{volume}{4}},
  \bibinfo{pages}{559} (\bibinfo{year}{1964}).

\bibitem[{\citenamefont{Bakhvalov}(1966)}]{Bakhvalov:1966}
\bibinfo{author}{\bibfnamefont{N.~S.} \bibnamefont{Bakhvalov}},
  \bibinfo{journal}{USSR Compt. Math. Math. Phys.}
  \textbf{\bibinfo{volume}{6}}, \bibinfo{pages}{861} (\bibinfo{year}{1966}).

\bibitem[{\citenamefont{Brandt}(1977)}]{BrandtA:1977}
\bibinfo{author}{\bibfnamefont{A.}~\bibnamefont{Brandt}},
  \bibinfo{journal}{Mathematics of Computation} \textbf{\bibinfo{volume}{31}},
  \bibinfo{pages}{333} (\bibinfo{year}{1977}).

\bibitem[{Bra()}]{Brandt77}
\bibinfo{note}{A. Brandt in {\it Multigrid Methods} edited by W.~Hackbusch and
  U.~Trottenberg (Springer--Verlag, Berlin, 1977)}.

\bibitem[{\citenamefont{Martin and Cartwright}(1996)}]{Martin:1996}
\bibinfo{author}{\bibfnamefont{D.}~\bibnamefont{Martin}} \bibnamefont{and}
  \bibinfo{author}{\bibfnamefont{K.}~\bibnamefont{Cartwright}},
  \bibinfo{journal}{unpublished}  (\bibinfo{year}{1996}), \bibinfo{note}{{\tt
  http://seesar.lbl.gov/anag/staff/martin/tar/ AMR.ps}}.

\bibitem[{\citenamefont{Briggs et~al.}(2000)\citenamefont{Briggs, Henson, and
  McCormick}}]{MGTutorial}
\bibinfo{author}{\bibfnamefont{W.~L.} \bibnamefont{Briggs}},
  \bibinfo{author}{\bibfnamefont{V.~E.} \bibnamefont{Henson}},
  \bibnamefont{and} \bibinfo{author}{\bibfnamefont{S.~F.}
  \bibnamefont{McCormick}}, \emph{\bibinfo{title}{A Multigrid Tutorial}}
  (\bibinfo{publisher}{Society for Industrial and Applied Mathematics},
  \bibinfo{address}{Philadelphia}, \bibinfo{year}{2000}).

\bibitem[{\citenamefont{MacNeice et~al.}(2000)\citenamefont{MacNeice, Olson,
  Mobarry, deFainchtein, and Packer}}]{MacNeice00}
\bibinfo{author}{\bibfnamefont{P.}~\bibnamefont{MacNeice}},
  \bibinfo{author}{\bibfnamefont{K.~M.} \bibnamefont{Olson}},
  \bibinfo{author}{\bibfnamefont{C.}~\bibnamefont{Mobarry}},
  \bibinfo{author}{\bibfnamefont{R.}~\bibnamefont{deFainchtein}},
  \bibnamefont{and} \bibinfo{author}{\bibfnamefont{C.}~\bibnamefont{Packer}},
  \bibinfo{journal}{Computer Physics Comm.} \textbf{\bibinfo{volume}{126}},
  \bibinfo{pages}{330} (\bibinfo{year}{2000}).

\bibitem[{par()}]{parameshmanual}
\bibinfo{note}{{\tt http://ct.gsfc.nasa.gov/paramesh/Users\_manual/ amr.html}}.

\bibitem[{\citenamefont{Brown and Lowe}(2004)}]{Brown:2004km}
\bibinfo{author}{\bibfnamefont{J.~D.} \bibnamefont{Brown}} \bibnamefont{and}
  \bibinfo{author}{\bibfnamefont{L.~L.} \bibnamefont{Lowe}},
  \bibinfo{journal}{Phys. Rev.} \textbf{\bibinfo{volume}{D70}},
  \bibinfo{pages}{124014} (\bibinfo{year}{2004}), \eprint{gr-qc/0408089}.

\bibitem[{\citenamefont{McCormick and Thomas}(1986)}]{McCormickFAC:1986}
\bibinfo{author}{\bibfnamefont{S.~F.} \bibnamefont{McCormick}}
  \bibnamefont{and} \bibinfo{author}{\bibfnamefont{J.}~\bibnamefont{Thomas}},
  \bibinfo{journal}{Mathematics of Computation} \textbf{\bibinfo{volume}{46}},
  \bibinfo{pages}{439} (\bibinfo{year}{1986}).

\bibitem[{\citenamefont{Bai and Brandt}(1987)}]{BaiBrandt:1987}
\bibinfo{author}{\bibfnamefont{D.}~\bibnamefont{Bai}} \bibnamefont{and}
  \bibinfo{author}{\bibfnamefont{A.}~\bibnamefont{Brandt}},
  \bibinfo{journal}{J. Sci. Stat. Comput.} \textbf{\bibinfo{volume}{8}},
  \bibinfo{pages}{109} (\bibinfo{year}{1987}).

\bibitem[{\citenamefont{Chesshire and Henshaw}(1990)}]{Henshaw:1990}
\bibinfo{author}{\bibfnamefont{G.}~\bibnamefont{Chesshire}} \bibnamefont{and}
  \bibinfo{author}{\bibfnamefont{W.~D.} \bibnamefont{Henshaw}},
  \bibinfo{journal}{J. Compt. Phys.} \textbf{\bibinfo{volume}{90}},
  \bibinfo{pages}{1} (\bibinfo{year}{1990}).

\bibitem[{\citenamefont{Press et~al.}(1992)\citenamefont{Press, Teukolsky,
  Vetterling, and Flannery}}]{NumericalRecipes}
\bibinfo{author}{\bibfnamefont{W.~H.} \bibnamefont{Press}},
  \bibinfo{author}{\bibfnamefont{S.~A.} \bibnamefont{Teukolsky}},
  \bibinfo{author}{\bibfnamefont{W.~T.} \bibnamefont{Vetterling}},
  \bibnamefont{and} \bibinfo{author}{\bibfnamefont{B.~P.}
  \bibnamefont{Flannery}}, \emph{\bibinfo{title}{Numerical Recipes}}
  (\bibinfo{publisher}{Cambridge University Press},
  \bibinfo{address}{Cambridge}, \bibinfo{year}{1992}).

\bibitem[{lap()}]{lapack_website}
\bibinfo{note}{{\tt http://www.netlib.org/lapack/index.html}}.

\bibitem[{\citenamefont{Trottenberg~U.}(2001)}]{MultigridTextBook}
\bibinfo{author}{\bibfnamefont{S.~A.} \bibnamefont{Trottenberg~U.},
  \bibfnamefont{Oosterlee C.~W.}}, \emph{\bibinfo{title}{Multigrid}}
  (\bibinfo{publisher}{Academic Press}, \bibinfo{address}{London},
  \bibinfo{year}{2001}).

\bibitem[{\citenamefont{Brandt and Br{\"u}gmann}(1997)}]{Brandt:1997tf}
\bibinfo{author}{\bibfnamefont{S.}~\bibnamefont{Brandt}} \bibnamefont{and}
  \bibinfo{author}{\bibfnamefont{B.}~\bibnamefont{Br{\"u}gmann}},
  \bibinfo{journal}{Phys. Rev. Lett.} \textbf{\bibinfo{volume}{78}},
  \bibinfo{pages}{3606} (\bibinfo{year}{1997}), \eprint{gr-qc/9703066}.

\bibitem[{\citenamefont{Bowen and York}(1980)}]{Bowen:1980yu}
\bibinfo{author}{\bibfnamefont{J.~M.} \bibnamefont{Bowen}} \bibnamefont{and}
  \bibinfo{author}{\bibfnamefont{J.}~\bibnamefont{York},
  \bibfnamefont{James~W.}}, \bibinfo{journal}{Phys. Rev.}
  \textbf{\bibinfo{volume}{D21}}, \bibinfo{pages}{2047} (\bibinfo{year}{1980}).

\end{thebibliography}
\end{document}